\documentclass[conference]{IEEEtran}
\IEEEoverridecommandlockouts

\usepackage{cite}
\usepackage{url}
\usepackage{multirow}
\usepackage{amsmath,amssymb,amsfonts}
\usepackage{algorithmic}
\usepackage{graphicx}
\usepackage{textcomp}
\usepackage{xcolor}
\def\BibTeX{{\rm B\kern-.05em{\sc i\kern-.025em b}\kern-.08em
    T\kern-.1667em\lower.7ex\hbox{E}\kern-.125emX}}
\begin{document}

\title{}

\title{FreeSVC: Towards Zero-shot Multilingual \\ Singing Voice Conversion\\
\thanks{This work has been fully/partially funded by the project Research and Development of Algorithms for Construction of Digital Human Technological Components supported by Advanced Knowledge Center in Immersive Technologies (AKCIT), with financial resources from the PPI IoT/Manufatura 4.0 / PPI HardwareBR of the MCTI grant number 057/2023, signed with EMBRAPII/}
}

\author{\IEEEauthorblockN{1\textsuperscript{st} Alef Iury Ferreira}
\IEEEauthorblockA{\textit{AKCIT Federal University of Goiás} \\
Goiânia, Goiás, Brazil \\
alef\_iury\_c.c@discente.ufg.br}
\and
\IEEEauthorblockN{2\textsuperscript{nd} Lucas Rafael Gris}
\IEEEauthorblockA{\textit{AKCIT Federal University of Goias} \\
Goiânia, Goiás, Brazil \\
lucas.gris@discente.ufg.br}
\and
\IEEEauthorblockN{3\textsuperscript{rd} Augusto da Rosa}
\IEEEauthorblockA{\textit{Federal Technological University of Paraná} \\
Medianeira, Paraná, Brazil \\
nosaveddataoz1@gmail.com}
\and
\IEEEauthorblockN{4\textsuperscript{th} Frederico Oliveira}
\IEEEauthorblockA{\textit{Federal University of Mato Grosso} \\
Cuiabá, Mato Grosso, Brazil \\
frederico.oliveira@ufmt.br}
\and
\IEEEauthorblockN{5\textsuperscript{th} Edresson Casanova} 
\IEEEauthorblockA{\textit{NVIDIA} \\
São Paulo, São Paulo, Brazil \\
ecasanova@nvidia.com}
\and
\IEEEauthorblockN{6\textsuperscript{th} Rafael Sousa}
\IEEEauthorblockA{\textit{Federal University of Mato Grosso} \\
Cuiabá, Mato Grosso, Brazil \\
rafael.sousa@ufmt.br}
\and 
\IEEEauthorblockN{7\textsuperscript{th} Arnaldo Junior }
\IEEEauthorblockA{\textit{São Paulo State University} \\
São Paulo, São Paulo, Brazil \\
arnaldo.candido@unesp.br}
\and
\IEEEauthorblockN{8\textsuperscript{th} Anderson Soares}
\IEEEauthorblockA{\textit{AKCIT Federal University of Goiás} \\
Goiânia, Goiás, Brazil \\
andersonsoares@ufg.br}
\and
\IEEEauthorblockN{9\textsuperscript{th} Arlindo Galvão Filho}
\IEEEauthorblockA{\textit{AKCIT Federal University of Goiás} \\
Goiânia, Goiás, Brazil \\
arlindo@inf.ufg.br}
}

\maketitle

\IEEEpubid{\begin{minipage}{\textwidth}\ \\[64pt] \centering \copyright~2025 IEEE. Personal use of this material is permitted. Permission from IEEE must be obtained for all other uses, in any current or future media, including reprinting/republishing this material for advertising or promotional purposes, creating new collective works, for resale or redistribution to servers or lists, or reuse of any copyrighted component of this work in other works.\end{minipage}}

\begin{abstract}


This work presents FreeSVC, a promising multilingual singing voice conversion approach that leverages an enhanced VITS model with Speaker-invariant Clustering (SPIN) for better content representation and the State-of-the-Art (SOTA) speaker encoder ECAPA2. FreeSVC incorporates trainable language embeddings to handle multiple languages and employs an advanced speaker encoder to disentangle speaker characteristics from linguistic content. Designed for zero-shot learning, FreeSVC enables cross-lingual singing voice conversion without extensive language-specific training. We demonstrate that a multilingual content extractor is crucial for optimal cross-language conversion. Our source code and models are publicly available\footnote{\url{https://github.com/freds0/free-svc}}.

\end{abstract}


\begin{IEEEkeywords}
Singing Voice Conversion, Synthesis of Singing Voices, Cross-lingual and multilingual aspects in speech synthesis.
\end{IEEEkeywords}

\section{Introduction}

Voice conversion (VC) is a technique that converts the voice of a source speaker to a target style, such as speaker identity~\cite{Li2023FreeVC}, prosody~\cite{wang2018style}, or emotion~\cite{zhou2021seen}, while preserving the linguistic content. Singing Voice Conversion (SVC) specifically converts the source singing voice to match the target speaker's voice, preserving the original lyrics and melody~\cite{Deng2020, Wang2021, Liu2021DiffSVC, Liu2021, Guo2022, Jayashankar2023, Zhou2023}.

SVC models often build upon VC advancements. Recent VITS-based~\cite{kim2021conditional} models exemplify this trend. In this context, YourTTS~\cite{casanova2022yourtts} extended VITS for zero-shot speech voice conversion, allowing multilingual training. FreeVC~\cite{Li2023FreeVC} further improved on this by integrating WavLM~\cite{chen2022wavlm} as the text encoder, enhancing speaker similarity for unseen speakers, though requiring speaker augmentation to manage speaker information leakage.

Inspired by these successes, VITS-based models were adapted for SVC, resulting in various open-source projects. A notable example is SoftVC VITS Singing Voice Conversion (so-vits-svc)\footnote{https://github.com/svc-develop-team/so-vits-svc}~\cite{Zhou2023}, which replaces the VITS text encoder with an SSL-based content encoder and conditions the model with pitch, utilizing a HuBERT-based~\cite{Hsu2021HuBERT} content encoder called ContentVec~\cite{qian2022contentvec}.




Despite advances in SVC, multilingual zero-shot capabilities remain underexplored. Existing zero-shot models~\cite{wu2022unified, jiang2023zero} are limited to single-language applications. This work investigates VITS-based architectures for zero-shot multilingual SVC, aiming to facilitate voice conversion across languages with minimal language-specific training, thus supporting low-resource languages. We propose the FreeSVC model, which integrates a state-of-the-art content encoder~\cite{spin}, a speaker encoder~\cite{Thienpondt2023ecapa2}, and trainable language embeddings, enabling cross-lingual SVC. Our contributions include:


\begin{itemize}
    \item Ablation studies exploring approaches to enhance zero-shot multilingual SVC;
    \item A model enabling cross-language singing voice conversions without extensive language-specific training.
\end{itemize}   

\section{FreeSVC Model}

\begin{figure*}[thb!]
  \centering
  \includegraphics[width=0.60\textwidth]{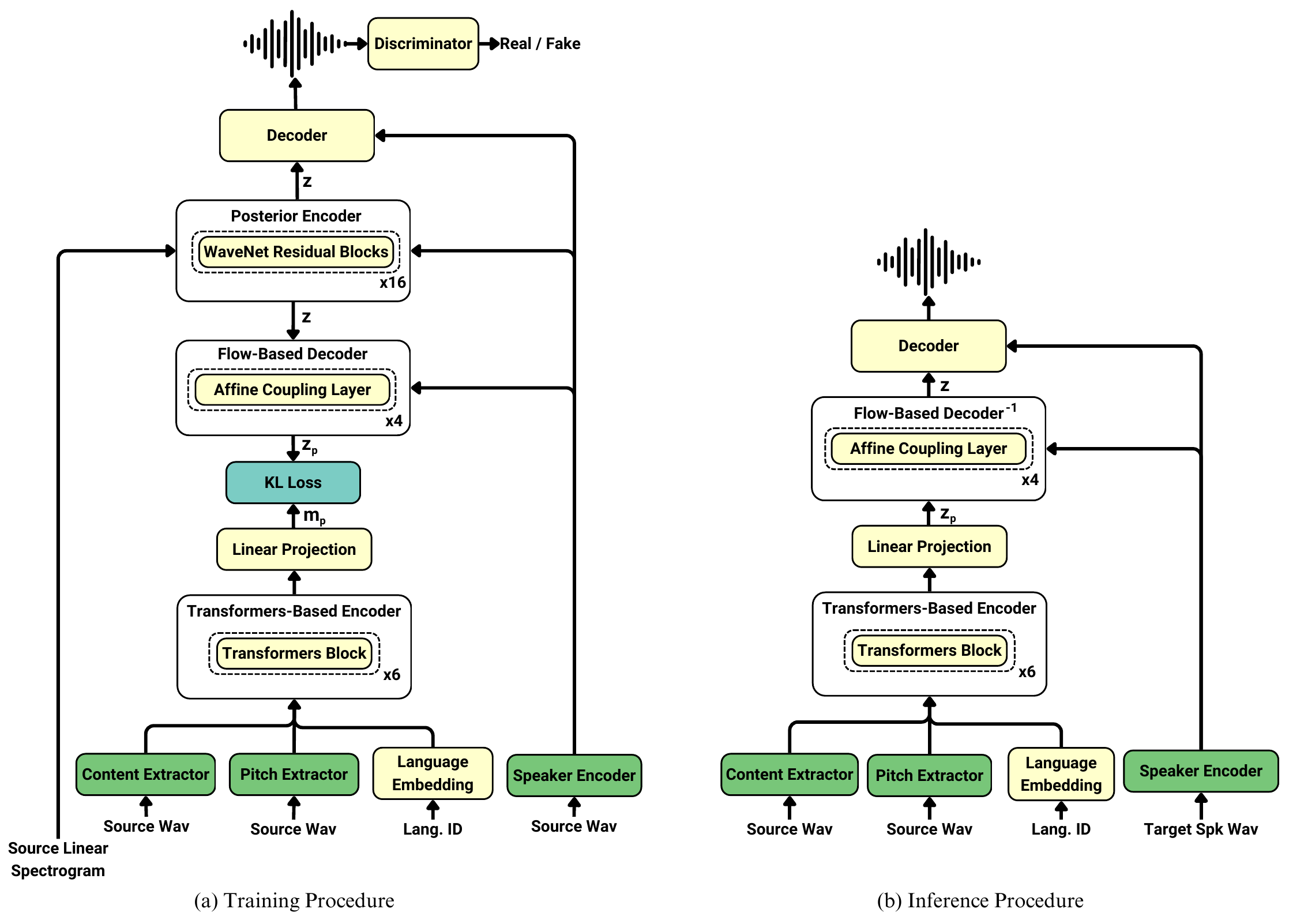}
  \caption{Comprehensive diagram of the FreeSVC model illustrating the (a) the training procedure, and (b) inference procedure.}
\label{fig:speech_production}
\end{figure*}

A comprehensive diagram of our proposed model is presented in Figure~\ref{fig:speech_production}.  Drawing inspiration from FreeVC, our method also originates from VITS~\cite{kim2021conditional}, incorporating training with Generative Adversarial Networks (GAN). Similar to FreeVC~\cite{Li2023FreeVC}, our initial encoder processes raw waveforms as input. The speaker embedding is obtained through a previously trained speaker encoder, while the content is derived from a Self-Supervised Learning (SSL) model. The architecture of the FreeSVC model is designed to address the challenges of multilingual SVC, harnessing the capabilities of VITS for end-to-end training with the addition of novel components to enhance content extraction, speaker encoding, and pitch extraction. 

A key component of our system is the content extractor, which is responsible for extracting content from the source audio for synthesis. Recent models~\cite{Wang2021,Jayashankar2023,Zhou2023} use HuBERT~\cite{Hsu2021HuBERT} to separate timbre from linguistic content, effectively capturing linguistic features in singing voices across multiple languages. A commonly used variant is ContentVec~\cite{qian2022contentvec}, a HuBERT derivative. In our study, we adopt SPIN~\cite{spin}, which enhances ContentVec by incorporating a vector quantization layer during training and fine-tuning the last two layers with the same loss objective. During inference, this layer is removed, making SPIN’s architecture to functionally align with ContentVec.



Another key component is the speaker encoder, which is used to extract a speaker representation, capturing the unique characteristics of a speaker's voice. For this task, we employ ECAPA2~\cite{Thienpondt2023ecapa2} as the speaker encoder to extract speaker/singer identity embeddings. ECAPA2's hybrid architecture combines 1D and 2D convolutions, optimizing it for capturing vocal characteristics, even in challenging scenarios like overlapping voices or short speech segments. We use official checkpoints\footnote{\url{https://huggingface.co/Jenthe/ECAPA2}}, trained on the VoxCeleb2 dataset~\cite{Chung18b}, ensuring robust and high-quality embeddings.



Finally, to preserve the notes in the singing, it is essential to model the pitch of the source audio. The RMVPE~\cite{wei2023rmvpe} was employed as a Pitch Extractor in our model, due to its robustness, even when dealing with audios that have background music. Typically, music source separation is required before pitch estimation, but this step can reduce accuracy depending on the separation quality. RMVPE addresses this by directly extracting vocal pitches from polyphonic music, leveraging deep U-Net \cite{unet} and GRU \cite{gru} networks to identify effective hidden features and accurately predict vocal pitches.

\section{Dataset}

We make use of a wide array of datasets covering both speech and singing, featuring various languages, hours of audio, and numbers of speakers. Table~\ref{tab:datasets} presents a comprehensive list of the datasets utilized in the experiments. For speech, datasets like AISHELL-1~\cite{Bu2017Aishell1} and AISHELL-3~\cite{Shi2021Aishell3} offer Mandarin audio, while CML-TTS~\cite{Oliveira2023CML} provides over 3,000 hours of speech across seven languages: Dutch, French, German, Italian, Portuguese, Polish, and Spanish. JVS~\cite{Takamichi2019JVS} contributes 30 hours of Japanese speech. HiFiTTS~\cite{Bakhturina21HiFiTTS} and LibriTTS-R~\cite{Koizumi2023librittsr} focus on English, with hundreds of hours of audio and numerous speakers. VCTK~\cite{Yamagishi2019VCTK} is notable for its diversity in English speech, making it particularly useful for voice cloning research.

\begin{table}[ht]
\setlength{\tabcolsep}{4.5pt} 
\renewcommand{\arraystretch}{1} 
\caption{Summary of the Datasets used for training.} 
\centering
\resizebox{0.40\textwidth}{!}{%
\begin{tabular}{c|c|c|c|c|c}
\hline
\textbf{Dataset} & \textbf{Hours} & \multicolumn{2}{|c|}{\textbf{Speakers}} & \textbf{Lang} & \textbf{Type} \\
         &  & F & M &  &  \\
\hline
AISHEL-1 \cite{Bu2017Aishell1} & 170h & 214 & 186  & CH & Speech \\
AISHEL-3 \cite{Shi2021Aishell3} & 85h & 176 & 42 & CH  & Speech \\
CML-TTS \cite{Oliveira2023CML} & 3.1k & 231 & 194  & 7 & Speech \\ 
HiFiTTS \cite{Bakhturina21HiFiTTS} & 292h & 6 & 4 & EN & Speech \\ 
JVS \cite{Takamichi2019JVS} & 30h & 51 & 49  & JP & Speech \\
LibriTTS-R \cite{Koizumi2023librittsr} & 585h & \multicolumn{2}{c|}{2,456} & EN & Speech \\ 
NUS (NHSS) \cite{Sharma20219NHSS} & 7h & 5 & 5  & EN & Both \\ 
OpenSinger \cite{Rongjie2021OpenSinger} & 50h & 41 & 25 & CH &  Singing \\ 
Opencpop \cite{Wang2022OPencpop} & 5h & 1 & - & CH &  Singing \\ 
\multirow{2}{*}{PopBuTFy \cite{Liu2022PopBuTFy} }& 10h & \multicolumn{2}{c|}{12} & CH & \multirow{2}{*}{Singing} \\ 
                                                 & 40h & \multicolumn{2}{c|}{22} & EN & \\ 
POPCS \cite{Liu2021PopCS} & 5h & 1 & - & CH & Singing \\ 
VCTK \cite{Yamagishi2019VCTK} & 44h & \multicolumn{2}{c|}{109} & EN & Speech \\ 
VocalSet \cite{Wilkins2018VocalSet} & 10h & 11  & 9  & - & Singing \\
\hline
\end{tabular}
}
\label{tab:datasets}
\end{table}

Singing datasets include NUS (NHSS)~\cite{Sharma20219NHSS} with 7 hours of English audio, also including speech, OpenSinger~\cite{Rongjie2021OpenSinger} and Opencpop~\cite{Wang2022OPencpop} offer Mandarin recordings, POPCS~\cite{Liu2021PopCS} contains 5 hours of Mandarin singing from one female artist. PopBuTFy~\cite{Liu2022PopBuTFy}  presents 50 hours of audio in both Mandarin and English, and VocalSet provides 10 hours of varied vocal techniques. 

Our dataset organization strategy is tailored to distinguish between in-domain (known) and out-of-domain (unknown) speakers to the model. For evaluation involving in-domain speakers, we include up to ten audio samples per speaker from the training data, or just one if recordings are limited. For tests with out-of-domain speakers, we use the designated test sections from the CML and LibriTTS datasets. For all other datasets, we ensure the inclusion of one male and one female speaker.



\section{Experiments}

We conducted experiments across four primary variations of speaker encoder and language conditioning. In each experiment, we employed the RMVPE pitch extractor alongside a consistent base model architecture. The baseline configuration implements the SoftVC (so-vits-svc) architecture illustrated in Figure~\ref{fig:speech_production}, utilizing an English pretrained ContentVec model\footnote{\url{https://huggingface.co/lengyue233/content-vec-best}} for content extraction without language conditioning. We selected this baseline as it represents the current state-of-the-art in singing voice conversion, with demonstrated success in both research and practical applications, thus providing a strong foundation for evaluating our proposed improvements. From this foundation, we evaluated three variant configurations: (1) the baseline enhanced with language conditioning (Lang. Emb.), (2) the baseline incorporating our trained version of SPIN for content extraction (SPIN), and (3) the complete FreeSVC model integrating both language conditioning and the SPIN model (SPIN + Lang. Emb.).

\subsection{Training Strategy}

We trained a HuBERT-based model using the SPIN method on CML-TTS and LibriTTS datasets for 3 epochs to obtain multilingual speech representations. The model used a batch size of 32, a cluster size of 2,048, and a dimensionality of 256 for the vector quantization layer. Other settings were kept the same as those described in~\cite{spin}.


For FreeSVC, we used the weights from the original FreeVC model. All configurations were fine-tuned using audio sampled at 24 kHz, except for the content extractor and speaker embedding model, which used a sampling rate of 16 kHz. We used all datasets listed in Table~\ref{tab:datasets} and trained for 225k steps (approximately 15 epochs) on a single A100 GPU with 80 GB of VRAM. We optimize with the Adam Optimizer~\cite{kingma2014Adam} with a learning rate set to $2e-4$, ($\beta_{1}$ = 0.8, $\beta_{2}=0.99$) and $\epsilon=1e-9$, using a batch size of 128 samples. To ensure fair distribution of both major and minor languages and speakers throughout the fine-tuning process, we employed weighted random sampling. This technique adjusts each sample's weight inversely to its frequency within the dataset, promoting equitable representation in each batch. Additionally, the linear spectrogram used in the training phase passed to the Posterior Encoder is extracted from the raw waveform with a window length of 1,280 samples and hop size of 320 samples.

\subsection{Evaluation Metrics}

We conducted both subjective and objective evaluations. For subjective evaluation, 46 participants assessed the naturalness of converted samples (combining both speech and singing in the evaluation)  using a 5-point Likert scale (1-bad, 2-poor, 3-fair, 4-good, 5-excellent) to calculate Mean Opinion Scores (MOS). For objective evaluation, we used Word Error Rate (WER) and Character Error Rate (CER), calculating the error between source and converted samples with Massively Multilingual Speech (MMS)~\cite{mms} Automatic Speech Recognition model. Zero-shot conversion quality was evaluated using cosine similarity between known and unknown speaker embeddings.

To compare approaches against the baseline, we employed bootstrapping~\cite{efron1994} to estimate confidence intervals, generating 1,000 bootstrap samples. We constructed $95\%$ confidence intervals using the $2.5$th and $97.5$th percentiles, utilizing the tool developed by Ferrer and Riera~\cite{ferrer2024Confidence}.

\section{Results}


In this section, we present our results through objective and subjective evaluations.

\subsection{Objective Evaluations}


In order to evaluate the zero-shot capabilities of our model, we calculated average speaker embeddings for each speaker in the designated test sets and corresponding generated audio produced by each model, as shown in Table~\ref{tab:spk}. No significant differences were observed among the models. This aligns with our expectations, since neither the content extractor nor the Lang. Emb components are supposed to impact the model's zero-shot capability unless there were inadvertent leaks of speaker information from these components. As anticipated, known speakers exhibited higher scores in the context of speech, while performance decreased for singing or unknown speakers. Notably, the values for unknown and known singing speakers were similar, suggesting robust performance across varying speaker characteristics.

\begin{table}[h]  

\caption{Comparison of average speaker embeddings: dataset vs. generated test audios, with standard deviations ($\pm$std). ($^{**}$) indicates significant difference from ContentVec baseline.}

\centering
\resizebox{0.37\textwidth}{!}{%
\begin{tabular}{l|cc|c}
\hline
\multicolumn{1}{c|}{\multirow{2}{*}{\textbf{Model}}} & \multicolumn{2}{c|}{\textbf{Known Speakers}}                                                                                               & \multirow{2}{*}{\textbf{\begin{tabular}[c]{@{}c@{}}Unknown \\ Speakers\end{tabular}}} \\ \cline{2-3}
\multicolumn{1}{c|}{}                                & \multicolumn{1}{l|}{\textbf{Singing}}                                          & \multicolumn{1}{l|}{\textbf{Speech}}                      &                                                                                       \\ \hline
ContentVec                                            & \multicolumn{1}{c|}{\begin{tabular}[c]{@{}c@{}}0.192$\pm$\\ 0.015\end{tabular}} & \begin{tabular}[c]{@{}c@{}}0.285$\pm$\\ 0.022\end{tabular} & \begin{tabular}[c]{@{}c@{}}0.185$\pm$\\ 0.016\end{tabular}                             \\ \hline
Lang. Emb.                                            & \multicolumn{1}{c|}{\begin{tabular}[c]{@{}c@{}}0.177$\pm$\\ 0.013$^{**}$\end{tabular}} & \begin{tabular}[c]{@{}c@{}}0.277$\pm$\\ 0.021$^{**}$\end{tabular} & \begin{tabular}[c]{@{}c@{}}0.182$\pm$\\ 0.013\end{tabular}                             \\ \hline
SPIN                                                  & \multicolumn{1}{c|}{\begin{tabular}[c]{@{}c@{}}0.172$\pm$\\ 0.015$^{**}$\end{tabular}} & \begin{tabular}[c]{@{}c@{}}0.268$\pm$\\ 0.024$^{**}$\end{tabular} & \begin{tabular}[c]{@{}c@{}}0.189$\pm$\\ 0.016\end{tabular}                             \\ \hline
SPIN + Lang. Emb.                                     & \multicolumn{1}{c|}{\begin{tabular}[c]{@{}c@{}}0.181$\pm$\\ 0.015$^{**}$\end{tabular}} & \begin{tabular}[c]{@{}c@{}}0.284$\pm$\\ 0.026\end{tabular} & \begin{tabular}[c]{@{}c@{}}0.197$\pm$\\ 0.016$^{**}$\end{tabular}                             \\ \hline

\end{tabular}\label{tab:spk}
}
\end{table}

Furthermore, we assessed content preservation by transcribing the generated audio samples using the MMS transcription model. Both our SVC model and MMS introduce cumulative errors, resulting in elevated Word Error Rates (WER) and Character Error Rates (CER), as shown in the \textbf{GT Text} column of Table~\ref{tab:speech-content-eval}. To mitigate evaluation bias, we also transcribed ground-truth audios with MMS. We then computed the WER and CER between the ground truth and synthesized audios' transcriptions, as presented in the \textbf{Transcription} column. Table~\ref{tab:speech-content-eval} shows consistent improvements in these metrics, indicating that FreeSVC achieves enhanced preservation of linguistic content. These enhancements are attributed to SPIN's more accurate content representation, which clarifies linguistic elements and reduces conversion errors.

\begin{table}[h]
\centering


\caption{Word Error Rate (WER) and Character Error Rate (CER) for generated audios. \textbf{GT Text}: Evaluation between ground truth transcriptions and model-generated transcriptions. \textbf{Transcription}: Evaluation between MMS transcriptions of ground truth audios and model-generated transcriptions.}

\centering
\resizebox{0.38\textwidth}{!}{%
\begin{tabular}{l|r r|r r}
\hline
& \multicolumn{2}{c|}{\textbf{GT Text}} & \multicolumn{2}{c}{\textbf{Transcription}} \\
\hline
\textbf{Model} & \textbf{WER}
 & \textbf{CER} & \textbf{WER} & \textbf{CER} \\
\hline
 ContentVec & 49.80 & 27.97 & 29.60 & 13.97 \\
 Lang. Emb. & 49.66 & 27.70 & 29.42 & 13.76 \\
 SPIN & 44.33 & 24.65 & 22.49 & 10.16 \\
 SPIN + Lang. Emb. & \textbf{44.27} & \textbf{24.58} & \textbf{22.39} & \textbf{10.07} \\
\hline
\end{tabular}}
\label{tab:speech-content-eval}
\end{table}

Additionally, to assess the prosody of singing voice conversion, we utilized the F0PPC (F0 Pearson Correlation Coefficient) after applying dynamic time warping (DTW) to align the original and generated audio. The integration of SPIN and Language Embedding significantly enhances singing voice conversion, as evidenced by the highest F0PPC scores for both known and unknown speakers in Table~\ref{tab:f0-eval}. This combination proves superior in delivering more natural and accurate prosody in singing voices.

\begin{table}[ht]
\setlength{\tabcolsep}{4.5pt} 
\renewcommand{\arraystretch}{1} 
\centering

\caption{F0PPC comparison between dataset and generated test set audios, with standard deviations ($\pm$std). ($^{**}$) denotes significant difference from ContentVec baseline.}

\centering
\resizebox{0.42\textwidth}{!}{%
\begin{tabular}{l|c|c}
\hline
\textbf{Model} & {\textbf{Known Speakers}} & {\textbf{Unknown Speakers}} \\
\hline
 ContentVec & 0.931$\pm$0.131 & 0.913$\pm$0.14 \\
 Lang. Emb. & 0.937 $\pm$0.123$^{**}$ & 0.913$\pm$0.14\\
 SPIN & 0.940$\pm$0.114$^{**}$ & 0.926$\pm$0.131$^{**}$\\
 SPIN + Lang. Emb. & \textbf{0.951$\pm$0.098$^{**}$} & \textbf{0.935$\pm$0.117$^{**}$}\\
\hline
\end{tabular}}
\label{tab:f0-eval}
\end{table}

\subsection{Subjective Evaluation}

\begin{table}[ht]
\setlength{\tabcolsep}{4.5pt} 
\renewcommand{\arraystretch}{1} 
\centering

\caption{Mean Opinion Score (MOS) evaluation of zero-shot conversion across target languages, with standard deviations ($\pm$std). \textbf{Other} denotes non-linguistic samples from VocalSet; \textbf{English} and \textbf{Chinese} indicate reference languages.}

\centering
\resizebox{0.40\textwidth}{!}{%
\begin{tabular}{l|c|c|c}
\hline
\textbf{Model} & {\textbf{English}} & {\textbf{Chinese}} & {\textbf{Other}} \\
\hline
 Ground Truth & 4.76$\pm$0.59 & 4.62$\pm$0.79 & 3.84$\pm$1.50\\
 \hline
 ContentVec & 3.02$\pm$1.08 & 2.89$\pm$1.03 & 1.32$\pm$0.63\\
 Lang. Emb. & 3.08$\pm$1.01 & 2.95$\pm$1.00 & 1.39$\pm$0.74 \\
 SPIN & 3.15$\pm$1.06 & 3.02$\pm$1.03 & 1.45$\pm$0.70\\
 SPIN + Lang. Emb. & \textbf{3.16$\pm$1.03} & \textbf{3.14$\pm$0.99} & \textbf{1.49$\pm$0.73}\\
\hline
\end{tabular}}
\label{tab:mos-eval}
\end{table}


The MOS results in Table \ref{tab:mos-eval} demonstrate that FreeSVC outperforms the previous baseline in zero-shot conversion. The integration of SPIN and the introduction of a language encoder have contributed to more natural and intelligible speech output across multiple languages, indicating the effectiveness of these modifications. The addition of language embedding slightly improves quality, supporting the hypothesis that better content representation and multilingual input handling are crucial for enhancing the subjective experience of converted voices.


Table~\ref{tab:mos-lingual} presents the MOS results aggregated by intra-lingual and cross-lingual voice conversion samples. The data indicates that language embedding does not yield significant improvements in intra-lingual settings; however, it enhances performance in both cross-lingual scenarios. Notably, SPIN demonstrates a modest improvement across both tasks, which may be attributed to the multilingual fine-tuning process employed.

\begin{table}[h] 
\setlength{\tabcolsep}{4.5pt}
\renewcommand{\arraystretch}{1}
\centering
\caption{Zero-Shot evaluation of intra-lingual and cross-lingual conversion of the MOS ratings and their respective standard deviations ($\pm$std).}
\resizebox{0.45\textwidth}{!}{%
\begin{tabular}{l|cc|cc}
\hline
\multirow{2}{*}{\textbf{Model}} & \multicolumn{2}{c|}{\textbf{Intra-lingual}}              & \multicolumn{2}{c}{\textbf{Cross-lingual}}               \\ \cline{2-5} 
                                & \multicolumn{1}{c|}{\textbf{English}} & \textbf{Chinese} & \multicolumn{1}{c|}{\textbf{English}} & \textbf{Chinese} \\ \hline
Ground Truth                    & \multicolumn{1}{c|}{4.82$\pm$0.47}             & 4.82$\pm$0.38             & \multicolumn{1}{c|}{4.75$\pm$0.61}             & 4.61$\pm$0.80 \\ \hline
ContentVec                      & \multicolumn{1}{c|}{3.39$\pm$1.11}             & 3.0$\pm$0.97              & \multicolumn{1}{c|}{2.92$\pm$1.05}             & 2.88$\pm$1.03        \\
Lang. Emb.                      & \multicolumn{1}{c|}{3.36$\pm$1.33}             & 2.94$\pm$0.87             & \multicolumn{1}{c|}{3.02$\pm$0.91}             & 2.95$\pm$1.00            \\
SPIN                            & \multicolumn{1}{c|}{3.38$\pm$1.07}             & \textbf{3.13$\pm$0.72}    & \multicolumn{1}{c|}{\textbf{3.09$\pm$1.05}}    & 3.01$\pm$1.04             \\
SPIN + Lang. Emb.               & \multicolumn{1}{c|}{\textbf{3.58$\pm$1.25}}    & 3.07$\pm$0.88             & \multicolumn{1}{c|}{3.06$\pm$0.94}             & \textbf{3.14$\pm$0.99}    \\ \hline
\end{tabular}}
\label{tab:mos-lingual}
\end{table}

\section{Conclusions}



In conclusion, this study has highlighted the significant challenges inherent in zero-shot and multilingual singing voice conversion, exacerbated by cross-lingual variations and limited data availability. To address these issues, we proposed a FreeVC model variant incorporating language and speaker embedding conditioning, trained on both singing and speech data across multiple languages. The fine-tuning of a HuBERT-based model demonstrated that a multilingual content extractor is crucial for optimal performance, serving as a key factor in managing diverse linguistic inputs.


Future research will focus on improving model performance by developing a speaker encoder trained on singing data and refining the content extractor, aiming to enhance accuracy and quality in multilingual singing voice conversion.

\bibliographystyle{IEEEtran}
\bibliography{mybib}

\end{document}